\documentclass[prl,aps,nofootinbib,twocolumn]{revtex4}
\usepackage{graphicx} 
\usepackage{dcolumn}  
\usepackage{bm}       
\usepackage{amsmath}
\usepackage{amssymb}
\usepackage{epsfig}

\newcommand{\es}{\mathbf{\hat{e}}_s} 
\newcommand{\eppm}{\mathbf{\hat{e}}_p^\pm} 
\newcommand{\espm}{\mathbf{\hat{e}}_s^\pm} 
\newcommand{\vpm}{\mathbf{v}^\pm} 
\newcommand{\kx}{k_x}
\newcommand{\ky}{k_y}
\newcommand{\kz}{k_z}
\newcommand{\px}{p_x}
\newcommand{\py}{p_y}
\newcommand{\pz}{p_z}
\newcommand{\mx}{m_x}
\newcommand{\my}{m_y}
\newcommand{\mz}{m_z}

\begin{document}
\title{Amplitude and phase control of guided modes excitation from a single dipole source: engineering far- and near-field directionality}
 \author{Michela F. Picardi}
 \affiliation{Department of Physics and London Centre for Nanotechnology, King’s College London, Strand, London WC2R 2LS, UK}
\author{Anatoly V. Zayats}
\affiliation{Department of Physics and London Centre for Nanotechnology, King’s College London, Strand, London WC2R 2LS, UK}
 \author{Francisco J. Rodr\'{i}guez-Fortu\~{n}o}
  \email[Corresponding author: ]{francisco.rodriguez\_fortuno@kcl.ac.uk}
 \affiliation{Department of Physics and London Centre for Nanotechnology, King’s College London, Strand, London WC2R 2LS, UK}


\begin{abstract}
The design of far-field radiation diagrams from combined electric and magnetic dipolar sources has recently found applications in nanophotonic metasurfaces that realize tailored reflection and refraction. Such dipolar sources also exhibit important near-field evanescent coupling properties with applications in polarimetry and quantum optics. Here we introduce a rigorous theoretical framework for engineering the angular spectra encompassing both far- and near-fields of electric and magnetic sources and develop a unified description of both free space and guided mode directional radiation. The approach uses the full parametric space of six complex-valued components of magnetic and electric dipoles in order to engineer constructive or destructive near-field interference. Such dipolar sources can be realized with dielectric or plasmonic nanoparticles. We show how a single dipolar source can be designed to achieve the selective coupling to multiple waveguide modes and far-field simultaneously with a desired amplitude, phase, and direction.
\end{abstract}

\maketitle
Nanophotonic components provide strong light-matter interactions, important for numerous applications, and the ability to efficiently control light with high bandwidth and energy efficiency \cite{atabaki2018integrating}. A standing challenge is the ultrafast switching and routing of light in miniaturized environments, preferably via an all-optical process. This can be achieved using the polarization of light \cite{Bliokh2015spinorbitreview}. In this context, dipolar scatterers with phased and balanced combinations of electric and magnetic moments have recently attracted significant theoretical and experimental attention due to the interesting, asymmetric, angular scattering properties of Huygens sources and Kerker-like conditions for far-field directionality and non-reflecting metasurfaces \cite{fu2013directional,Staude2013tailoring,liu2017huygens,paniagua2016generalized,Alaee2015generalized,coenen2014directional,Rolly2012boosting,Hancu2014multipolar}. Superposition of electric and magnetic dipoles in a single source can be realized experimentally with high-index dielectric nanoparticles in the Rayleigh regime \cite{fu2013directional, evlyukhin2012demonstration,kuznetsov2016optically,permyakov2015probing,wozniak2015selective,zywietz2014laser,picardi2019experimental} or using plasmonic scatterers with electric and magnetic responses \cite{Alaee2015generalized,coenen2014directional,Hancu2014multipolar}.

Beyond far-field directionality of dipolar sources, remarkable physics lies also in the near-field \cite{picardi2018not} where dipolar sources can be used for directional excitation of waveguided modes, relevant to integrated photonics. The routing of guided modes using subwavelength scatterers and emitters such as quantum dots or atoms, all of which can be modelled as dipolar sources, has been initially studied based on circularly polarized dipoles \cite{rodriguez2013near,petersen2014chiral,mitsch2014quantum,kapitanova2014photonic,neugebauer2014polarization,OConnor2014spin,coles2016chirality,young2015polarization,luxmoore2013optical}, and can be related to the spin-momentum locking characteristic of guided modes \cite{bliokh2015quantum,slobozhanyuk2019near,vanmechelen2016universal}. This effect has recently been generalized to dipole polarizations beyond circular, in which both the electric and magnetic dipole moments are important \cite{picardi2018janus,picardi2019experimental}. It was shown how three elemental dipolar sources can achieve near-field directionality: the circular, Huygens and Janus dipoles, with the latter two consisting of combinations of electric and magnetic dipoles. 
These three dipoles are elemental in the sense that each of them is built from the superposition of only two different dipole components, which is the minimum necessary to achieve destructive interference between them \cite{picardi2018not}. However, a general combination of one electric and one magnetic dipole has in fact 6 complex-valued components to work with $\mathbf{p}=(\px,\py,\pz)$ and $\mathbf{m}=(\mx,\my,\mz)$, each having an arbitrary amplitude and phase relative to the others. Thus, the parametric space of possibilities for near-field interference from general dipolar sources is much broader than previously realised.

In this work, we developed a generalised theory for far-field and near-field  dipolar directionality considering the entire three-dimensional polarization parameter space of both electric and magnetic components. 
With these uncovered new degrees of freedom, we can design destructive or constructive interference conditions for multiple modes simultaneously. For example, in a multimode waveguide, the directionality of individual modes can be controlled independently. We also show that both amplitude and phase of each excited mode can be individually controlled. This allows the engineering of mode distribution in multimode waveguides as well as the radiation pattern in free space.\\ 

\begin{figure*}[ht]
\includegraphics[width=\linewidth]{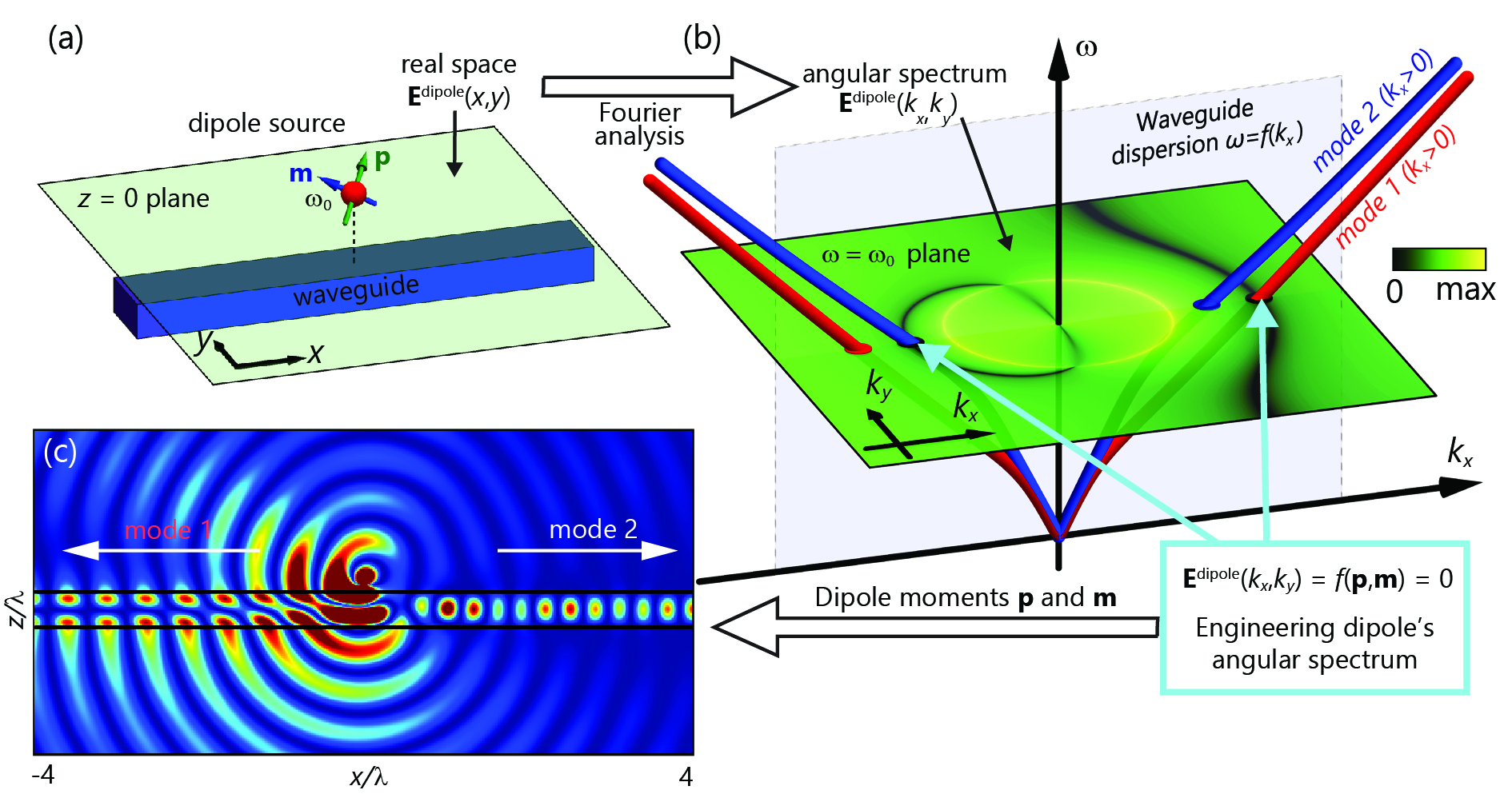}
\caption{Concept of dipolar angular spectra tailored to a nearby waveguide. (a) Schematic depiction of a dipole source placed near a waveguide. The $z=0$ plane, in which the waveguide lies, is the plane in which the fields of the dipole are calculated. (b)
The dipolar angular spectrum and the waveguide's dispersion relation plotted together in the frequency-momentum space. The dipole's angular spectrum can be engineered so that the spectral position of features of interest, such as zeros required for cancelling mode excitation, match the $k$-vector of guided modes. A multimode waveguide of thickness $t=0.6\lambda$ and refractive index $n=2$ surrounded by air, and dipole moments $p_x=-0.84i$, $p_z=-0.164$, and $m_y/c=1$ have been simulated. (c) Excitation of 2 different modes in opposite directions in the multimode waveguide obtained by cancelling required components of the dipolar fields as shown in (b). The plot corresponds to the dipole and waveguide parameters as in Fig. 3(c).}\label{fig:concept}
\end{figure*}

\textbf{Source angular spectrum.} The coupling of a dipole to a waveguide can be understood via Fermi's Golden Rule, where directional excitation of modes is interpreted as destructive interference between different dipolar components. The same phenomenon can be described using the angular spectrum representation of a source \cite{picardi2017unidirectional}, also called spatial spectrum or momentum representation. The electric field of a source can be expanded in momentum space via its angular spectrum  \cite{mandel1995optical,novotny2012principles,nieto2006scattering}:
\begin{equation}\label{eq:e_field}
    \mathbf{E}(x,y,z) = \iint \mathbf{E}(\kx,\ky, z)e^{i(\kx x +\ky y)}d\kx d\ky,
\end{equation}
where $x$, $y$, $z$ represent the spatial position, with $z$ being an arbitrary direction perpendicular to the plane in which we performed a 2D Fourier transform, and $\kx$ and $\ky$ represent the transverse wave-vector components. Since the field fulfills the wave-equation, the wave-vector components follow the relation $\mathbf{k} \cdot \mathbf{k} = k_x^2+k_y^2+ k_z^2 = k^2$, where $k=\omega \sqrt{\varepsilon\mu}$ is the wave-number of the medium and $\mathbf{k}$ is the wave-vector. The angular spectrum $\mathbf{E}(k_x,k_y)$ is defined at a given $z$-plane and represents a superposition of plane waves and evanescent waves. These waves can be propagated to other planes, to obtain the fields in the entirety of an homogeneous source-free space (via the transfer function $e^{\pm i k_z z}$) or at any point in a multi-slab structure via an adequate transfer function \cite{mandel1995optical}.

The design of the far-field of a source, i.e. its radiation diagram, is a common practice in antenna engineering \cite{balanis2005antenna}, and it can be interpreted as the design of the source's angular spectrum (Eq.~\ref{eq:e_field}) in the region of transverse momenta inside the light cone $(k_t<k\Rightarrow k_z\in \mathbb{R})$, where $k_t^2=k_x^2+k_y^2$ is the transverse wavevector, and each angular component $\mathbf{E}(k_x,k_y)$ corresponds to far-field radiation in a different direction in space. Such radiation diagram engineering is applied in nanophotonics for the design of asymmetric far-field scattering, reflection and transmission of photonic metasurfaces \cite{fu2013directional,Staude2013tailoring,liu2017huygens,paniagua2016generalized, Alaee2015generalized,coenen2014directional,Rolly2012boosting,Hancu2014multipolar}. In this work, we extend the engineering of the source angular spectrum beyond the light cone $(k_t>k)$ into a region of complex wave-vectors $(k_z\in \mathbb{I})$, to encompass the whole transverse momentum space. In this way, we include near field evanescent waves, responsible for the evanescent coupling of the source to guided modes in a nearby waveguide or surface modes on an adjacent surface.

\textbf{Simplified model of coupling to waveguided modes.} The wavevector components of guided modes are, by definition of their non-radiating guided nature, outside of the light cone. Therefore, our source can be designed to selectively overlap with the wave-vectors of the guided modes with engineered amplitudes and phases. For example, if a waveguide supports two modes at different wavelengths, then a source whose frequency spectrum lacks one of the two wavelengths will not excite the corresponding mode. This idea can be translated from the temporal to the spatial dimension.

In the same way as a time-invariant waveguide will conserve the frequency spectrum of the source, a translationally-invariant one will conserve the wavevector components parallel to the translation symmetry axes. A waveguide may support two modes with different wavevectors, so a source whose wavevector spectrum lacks one of the two components will not excite the corresponding mode. 
Consider the fields of a dipolar source in a given $z$-plane, $\mathbf{E}^{\text{dipole}} (x,y,t)$ as shown in Fig.~\ref{fig:concept}(a). We only look at the fields radiated by the dipole, its primary fields, not including the scattering by the waveguide. These fields can be Fourier-transformed to momentum-frequency space $\mathbf{E}^{\text{dipole}}(k_x,k_y,\omega)$ [Fig.\ref{fig:concept}(b)]. As the dipole is monochromatic, its spectrum is limited to a single frequency $\omega=\omega_0$ plane (although this could be generalised even further for time-varying sources). The dipole fields can now be represented together with the dispersion relation of the waveguide in the same momentum-frequency plot [Fig.\ref{fig:concept}(b)]: the intersection between the waveguide mode and the dipole fields is a requirement for matching the dipolar radiation to the modes. The dipole in this example has been engineered to exhibit zero amplitude exactly at certain points of intersection with the dispersion relation of the waveguide [Fig.~\ref{fig:concept}(b)]. Thus, this dipole will neither excite mode 1 to the right nor mode 2 to the left. As a result, each of the two modes will be excited, from the same dipolar source, into different directions of the waveguide. This is confirmed in real space [Fig.~\ref{fig:concept}(c)]. This description is exact in the case of a slab waveguide, in which guided modes have a single momentum value $(k_x,k_y)$ and transverse momentum is conserved. The model is not exact for one-dimensional waveguides such as that in the figure, because guided modes will have a spread of wave-vector $k_y$ values caused by the mode confinement along $y$, and the non- translational-invariance of the waveguide along $y$ will allow scattering along $k_y$. However, our simulations show that these effects are negligible for our purposes, and the model above becomes an excellent design principle even for one-dimensional waveguides.\\

\textbf{Complete model of coupling to waveguided modes.} When dealing with the angular spectrum of a dipole, the simple picture suggested by Fig.\,\ref{fig:concept} must be revised for two reasons. Firstly, the angular spectrum in Eq.\,\ref{eq:e_field} is a superposition of two spectra \cite{mandel1995optical}, $\mathbf{E}(k_x,k_y,z) = \mathbf{E}^{+}(k_x,k_y) e^{i k_z z} + \mathbf{E}^{-}(k_x,k_y) e^{-i k_z z}$, corresponding to waves propagating towards the positive and negative $z$ direction (we always define $k_z=\sqrt{k^2-k_t^2}$ as the positive root, with positive imaginary part if $k_t>k$). When dealing with a dipolar source, only the upwards propagating part $\mathbf{E}^+(k_x,k_y)$ will exist in the upper half space $z>0$, and only the downwards propagating part $\mathbf{E}^-(k_x,k_y)$ will exist in the lower half space $z<0$, since all waves must be propagating away from the source plane at $z=0$. Importantly, both spectra will be different in general. 

Secondly, each of these angular spectra $\mathbf{E}^{\pm} (k_x,k_y )$ is not a scalar function, as suggested by  Fig.~\ref{fig:concept}, but instead is a vector function. Despite being a three dimensional vector field, the divergence-free condition imposed by Maxwell's equations on each angular component $\mathbf{k}\cdot \mathbf{E}=0$ reduces by one the degrees of freedom, allowing the angular spectrum vector function to be written using a two-dimensional basis. A convenient choice of basis is to use $s$- and $p$-polarization: $\mathbf{E}^{\pm}(\kx,\ky) = E_s^{\pm}\espm + E_p^{\pm}\eppm$, where the two unit vectors $\espm = (1/k_t)(-\ky\hat{\mathbf{x}}+\kx\hat{\mathbf{y}})$ and $\eppm = (-1/k k_t)(\pm\kx\kz\hat{\mathbf{x}}\pm\ky\kz\hat{\mathbf{y}}+k_t\hat{\mathbf{z}})$ correspond to the azimuthal and polar angle unit vectors in spherical coordinates when describing far-fields $(k_t<k)$, but also accurately describe the evanescent wave polarizations (becoming complex vectors) when evaluated outside the light cone $(k_t>k)$. 

Evidently, the spectra for $s$- or $p$- polarized fields will couple to the corresponding $s$- or $p$-polarized waveguided modes in nearby waveguides, respectively. This is an exact statement in slabs, but the deviation from this rule for typical one-dimensional waveguides is negligible.

Therefore, a complete description of an arbitrary source requires different spectra to account for the waves with the two polarizations $s$- and $p$-, each defined on the half-spaces above and below the dipole, constituting \emph{four} different angular spectra in total. For example, Fig.~\ref{fig:concept} shows a waveguide supporting $p$-polarized modes placed below the dipole, so the relevant scalar angular spectrum is $E_p^- (k_x,k_y)$, which is the one shown.\\

\begin{figure}[ht]
\includegraphics[width=\linewidth]{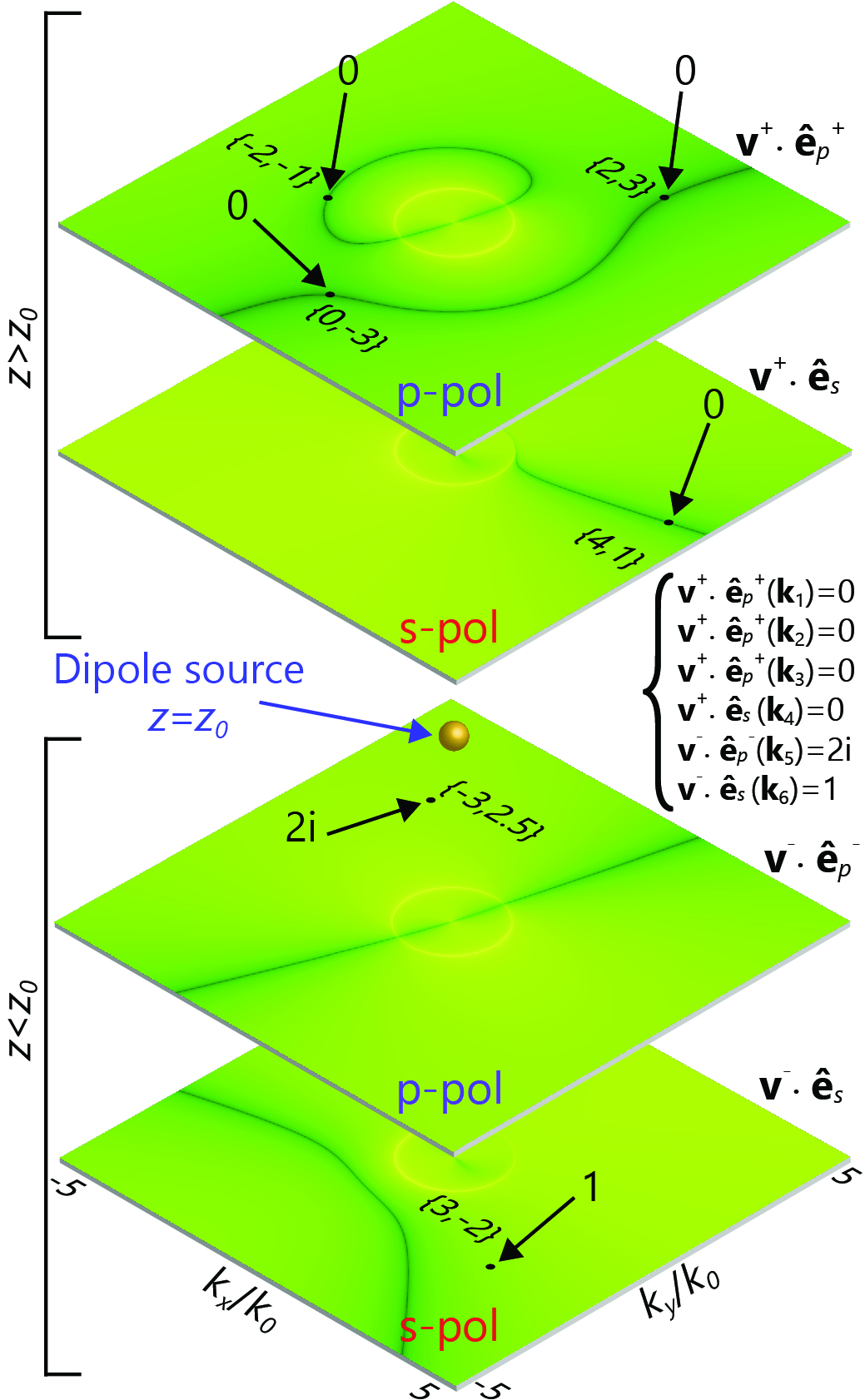}
\caption{Every dipolar source can be defined using four angular spectra, each specified by one of the two orthogonal polarization states, $s$ or $p$, and by the propagation direction along $z$, i.e. fields radiated above or below the source. The plot shows the four angular spectra of a single dipolar source, together with six chosen values of the electric field amplitude and phase (four of them being zero, and the other two being $1$ and $2i$) placed in the four planes at different locations in $k$-space (indicated with arrows). The dipole moment components required to achieve this combination were found to be $\mathbf{p}=(3.10, -1.00, 0.26i)$ and $\mathbf{m}/c=(-2.03i, -6.88i, 3.95)$.}\label{fig:planes}
\end{figure}

\textbf{Calculating dipolar angular spectra.} The angular spectra of dipolar sources can be obtained from Weyl's identity \cite{mandel1995optical}, and is typically expressed using dyadic Green functions \cite{novotny2012principles}. The spectra of dipoles can be re-written in an exact vector form describing the entire angular spectrum, including far- and near-fields as \cite{picardi2017unidirectional,picardi2018janus}:
\begin{equation}\label{eq:angular_spectrum}
    \mathbf{E}^{\pm}(\kx,\ky) = \frac{ik^2}{8\pi^2\varepsilon}\left[(\vpm\cdot\es)\es+(\vpm\cdot\eppm)\eppm\right],
\end{equation}
where $\vpm$ depends on the wave-vector and on both the electric $\mathbf{p}$ and magnetic $\mathbf{m}$ dipole moments of the source:
\begin{equation*}
    \vpm = \frac{1}{k_z}\left[\mathbf{p}-\left(\hat{\mathbf{e}}_k^{\pm}\times \frac{\mathbf{m}}{c}\right)\right],
\end{equation*}
where $\hat{\mathbf{e}}_k^{\pm} = \mathbf{k}^{\pm}/k$ and $\mathbf{k}^{\pm} = (k_x,k_y,\pm k_z)$. When  Eq.~\ref{eq:angular_spectrum} is applied to a given dipole $\mathbf{p}$ and $\mathbf{m}$, four spectra may be calculated and plotted, corresponding to $E^{+}_{p}$, $E^{+}_{s}$, $E^{-}_{p}$ and $E^{-}_{s}$. An example of these four spectra is shown in 
Fig.~\ref{fig:planes}.\\ 

\begin{figure}[ht]
\includegraphics[width=\linewidth]{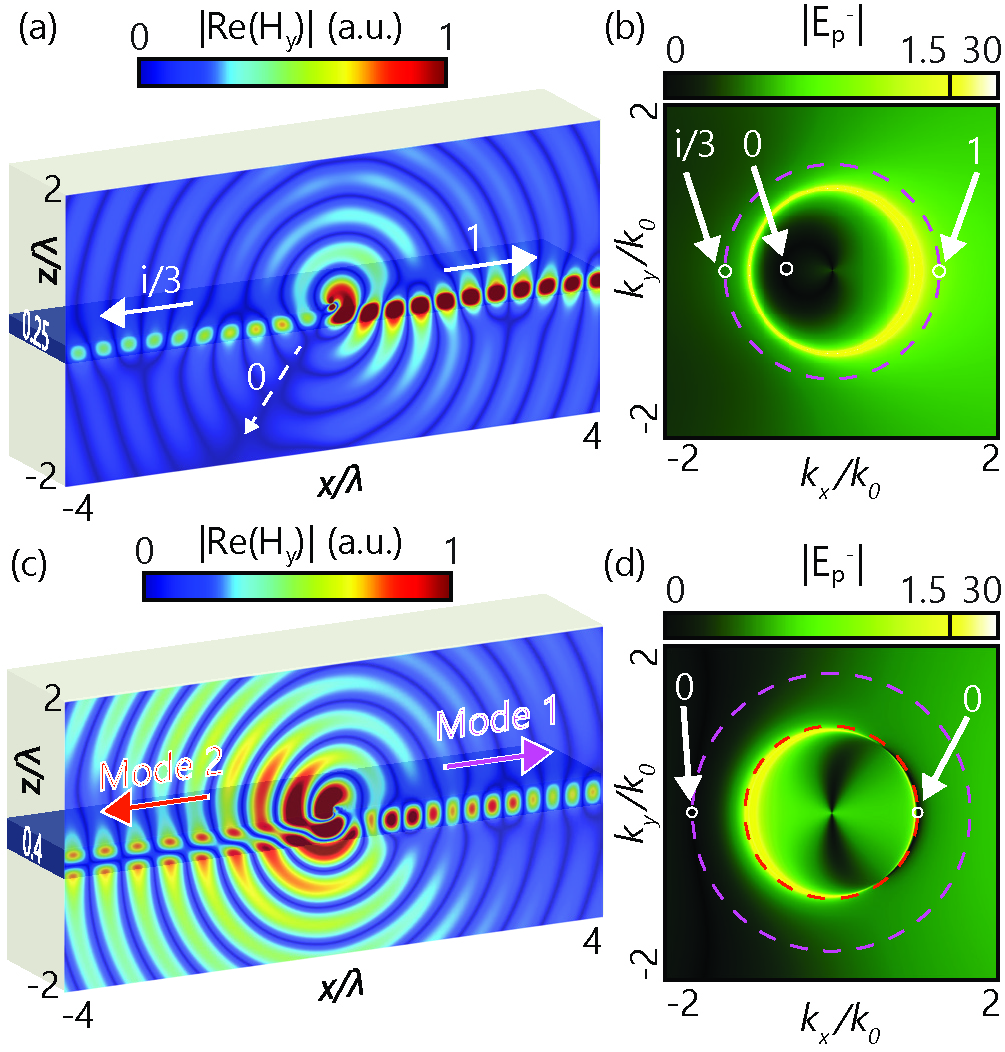}
\caption{Universal tunability: By suitably designing the amplitude and phase of the dipole at $k_y=0$ and $k_x = \pm k_{\mathrm{mode}}$, excitation of various  waveguided and free-space modes can be obtained. (a, b) The same mode is coupled in opposite directions with different amplitudes and $90^\circ$ out-of-phase, and a far-field direction of zero radiation is imposed below the source for an angle of $37^\circ$, [$\mathbf{p}=(1-0.03i,0,0.41-1.22i)$, $\mathbf{m}=(0,0.56-0.76i,0)$]. (c,d) Two different modes are coupled into opposite directions [$\mathbf{p}=(1i,0,-0.39)$,$\mathbf{m}/c=(0,-0.69,0)$]. (a, c) $\mathrm{Re}\left\lbrace H_y \right\rbrace$. (b, d) $E_p^-$ angular spectra of each dipole source. The radii of the dashed circles correspond to the wavevectors of the waveguide modes. The distance between the dipole and the waveguide is $0.05\,\lambda$. The refractive index of the waveguide is $n=2.2$ and the surrounding material is air. The thickness of the waveguide is $0.25\,\lambda$ in (a, b) and $0.4\,\lambda$ in (c,d) \label{fig:three_cases}}
\end{figure}

\begin{figure}[h!t]
\includegraphics[width=\linewidth]{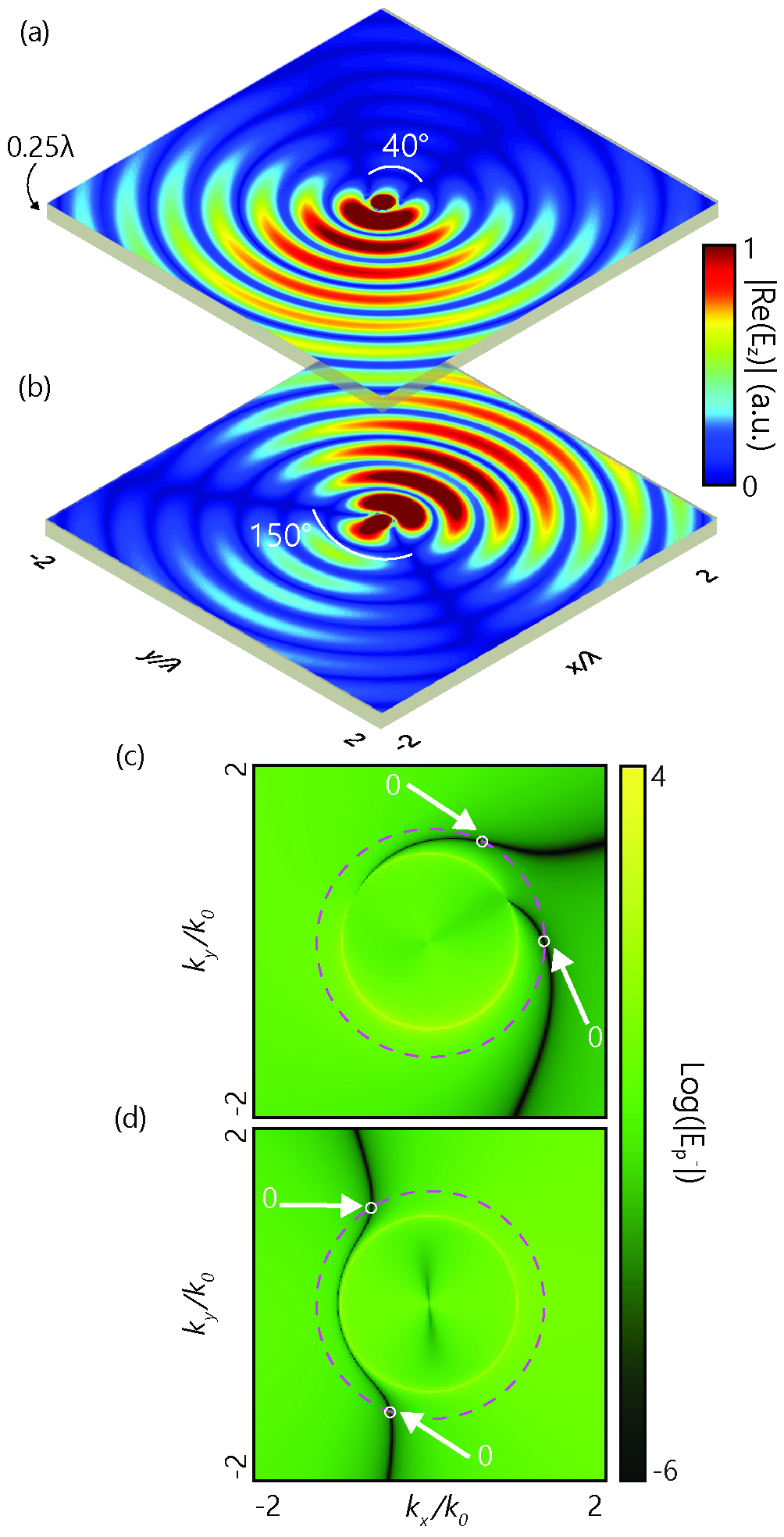}
\caption{Engineering the in-plane steering of the guided mode in a planar waveguide in arbitrary directions. (a, c) The near-field electric field amplitude is designed to have two zeros separated by $40^\circ$ [$\mathbf{p}=(1.56i,0.90i,1)$]. (b,d) The zeros are separated by $150^\circ$ [$\mathbf{p}=(3.30 i, 0, 2.11)$, $\mathbf{m}=(-1, 0, 0)$]]. The position of zeros controls the orientation and opening angle of the radiation pattern. (a, b) $\mathrm{Re}\left\lbrace E_z \right\rbrace$. (c, d) $E_p^-$ angular spectra of each dipole source. The radii of the dashed circles correspond to the wavevectors of the waveguide modes. The distance between the dipole and the waveguide is $0.05\,\lambda$, while the thickness of the waveguide is $0.25\,\lambda$. The refractive index of the waveguide is $n=2.2$ and the surrounding material is air.\label{fig:planar_steering}}
\end{figure}

\textbf{Engineering dipolar angular spectra.} So far, we have described the direct problem: given dipoles $\mathbf{p}=(p_x,p_y,p_z)$ and $\mathbf{m}=(m_x,m_y,m_z)$, we obtain the four relevant angular spectra which are a unique signature of this dipole via Eq.~\ref{eq:angular_spectrum}. The equation is linear, so coherent superposition of dipoles will result in the linear combination of their fields and spectra. Now, we want to solve the inverse problem. Our aim is to design a dipole whose angular spectra takes prescribed complex scalar values at designed points (specifying both amplitude \emph{and} phase of the spectrum at each point), distributed throughout the four scalar spectra, as shown for example in  Fig.~\ref{fig:planes}. Each condition is expressed by equating Eq.~\ref{eq:angular_spectrum} to the desired value. This forms a system of equations with 6 complex-valued unknowns $(p_x,p_y,p_z,m_x,m_y,m_z)$ and as many equations as specific conditions that we impose onto the angular spectra. 
Expressed in matrix notation, basic linear algebra provides us with the solution(s), as described in the Supplementary Materials. Since we have 6 complex degrees of freedom in our dipole, we can specify up to 6 complex amplitudes at distinct points in $k$-space, distributed throughout the four scalar spectra. An example of such 6 conditions is shown in Fig.~\ref{fig:planes}, and the corresponding dipole satisfying them is given in the caption. We provide an online dipole calculator \cite{picardi2019calculator} to retrieve the dipole moments that achieve user-specified amplitudes in arbitrary points on the angular spectra. If our conditions are all placing zeroes, then we are limited to 5, because we need one degree of freedom in the solution to act as an arbitrary scaling coefficient of the dipolar solution (requiring 6 zeroes simply yields the trivial solution $\mathbf{p}=\mathbf{m}=0$). Other limitations apply: for example, if we try to place all our conditions in the same polarization or in the same angular direction, we will be limited in the number of conditions that we can place, because only a subset of the 6 dipole components are involved in that specific polarization and/or direction.
Overall, with this technique we acquire enormous design power. We can freely decide, within the above limitations, the amplitude and phase of excitation of nearby waveguided modes, even in multimode waveguides. Two such examples are shown in Fig.~\ref{fig:three_cases} and confirmed by electromagnetic simulations. In Fig.~\ref{fig:three_cases}\,(a), the $p$-polarized mode of the waveguide is selectively excited in opposite directions with different amplitudes and phases, while at the same time a zero scattering is imposed in the far field in a specified direction. In Fig.~\ref{fig:three_cases}\,(b), the directionality of two different modes of the same bimodal waveguide is designed so that each of the modes propagates unidirectionally in different directions. Further, Fig.~\ref{fig:planar_steering} shows the flexibility of the method to engineer direction and opening angle of the mode in the plane of the slab waveguide, by placing zeroes at required angles in the radiation diagram. The design space is enormous if multiple waveguides with different orientations and modes of different polarizations are considered.

In conclusion, we have extended the concept of radiation diagram engineering to encompass the entire angular spectrum, using a remarkably simple algebraic method. We provide a simple-to-use online calculator for the engineering of dipolar angular spectra \cite{picardi2019calculator}. With this tool, we can design the outside of the light cone, for engineering of evanescent guided mode excitation, as well as the inside, engineering the far-field radiation diagrams and radiated polarizations. The concept presented here is very general and its application can be extended in many ways, for instance, to time-varying sources in which the frequency dimension can also play a role, to higher order multipoles to gain more degrees of freedom, or to collections of sources such as arrays, which will further modify the angular spectra. In the case of arrays, the angular spectra will be discretized into diffraction orders both inside and outside the lightcone, introducing the possibility of metasurfaces aimed at near-field engineering. With this work we demonstrate an enormous design power for the nanophotonic control of finely-tailored waveguide excitations via the three-dimensional electric and magnetic polarization of single sources. With recent advances in the synthesis of complex beams, the required fine tuning of exotic electric and magnetic field polarizations to electrically and magnetically polarize nanoparticles in three dimensions is now an experimental reality \cite{neugebauer2016polarization,neugebauer2014polarization,wozniak2015selective,Wei2016excitation}.

\section{Acknowledgement}
This work was supported by European Research Council Starting Grant ERC-2016-STG-714151-PSINFONI, EPSRC (UK) and ERC iCOMM project (789340). All the data that support this research are provided in full in the results section and Supplementary Materials.

\bibliographystyle{ieeetr}
\bibliography{Multiple_zeros_v4.bib}

\begin{thebibliography}{10}

\bibitem{atabaki2018integrating}
A.~H. Atabaki, S.~Moazeni, F.~Pavanello, H.~Gevorgyan, J.~Notaros, L.~Alloatti,
  M.~T. Wade, C.~Sun, S.~A. Kruger, H.~Meng, K.~Al~Qubaisi, I.~Wang, B.~Zhang,
  A.~Khilo, C.~V. Baiocco, M.~A. Popović, V.~M. Stojanović, and R.~J. Ram,
  ``{Integrating photonics with silicon nanoelectronics for the next generation
  of systems on a chip},'' {\em Nature}, vol.~556, no.~7701, p.~349, 2018.

\bibitem{Bliokh2015spinorbitreview}
K.~Y. Bliokh, F.~J. Rodr{\'{i}}guez-Fortu{\~{n}}o, F.~Nori, and A.~V. Zayats,
  ``{Spin–orbit interactions of light},'' {\em Nature Photonics}, vol.~9,
  pp.~796--808, dec 2015.

\bibitem{fu2013directional}
Y.~H. Fu, A.~I. Kuznetsov, A.~E. Miroshnichenko, Y.~F. Yu, and B.~Luk'yanchuk,
  ``{Directional visible light scattering by silicon nanoparticles},'' {\em
  Nature communications}, vol.~4, p.~1527, 2013.

\bibitem{Staude2013tailoring}
I.~Staude, A.~E. Miroshnichenko, M.~Decker, N.~T. Fofang, S.~Liu, E.~Gonzales,
  J.~Dominguez, T.~S. Luk, D.~N. Neshev, I.~Brener, and Y.~Kivshar,
  ``{Tailoring directional scattering through magnetic and electric resonances
  in subwavelength silicon nanodisks},'' {\em ACS Nano}, vol.~7,
  pp.~7824--7832, sep 2013.

\bibitem{liu2017huygens}
S.~Liu, A.~Vaskin, S.~Campione, O.~Wolf, M.~B. Sinclair, J.~Reno, G.~A. Keeler,
  I.~Staude, and I.~Brener, ``{Huygens' metasurfaces enabled by magnetic dipole
  resonance tuning in split dielectric nanoresonators},'' {\em Nano letters},
  vol.~17, no.~7, pp.~4297--4303, 2017.

\bibitem{paniagua2016generalized}
R.~Paniagua-Dom{\'{i}}nguez, Y.~F. Yu, A.~E. Miroshnichenko, L.~A. Krivitsky,
  Y.~H. Fu, V.~Valuckas, L.~Gonzaga, Y.~T. Toh, A.~Y.~S. Kay, B.~Luk'yanchuk,
  and A.~I. Kuznetsov, ``{Generalized Brewster effect in dielectric
  metasurfaces},'' {\em Nature Communications}, vol.~7, p.~10362, jan 2016.

\bibitem{Alaee2015generalized}
R.~Alaee, R.~Filter, D.~Lehr, F.~Lederer, and C.~Rockstuhl, ``{A generalized
  Kerker condition for highly directive nanoantennas},'' {\em Optics Letters},
  vol.~40, p.~2645, jun 2015.

\bibitem{coenen2014directional}
T.~Coenen, F.~{Bernal Arango}, A.~{Femius Koenderink}, and A.~Polman,
  ``{Directional emission from a single plasmonic scatterer},'' {\em Nature
  Communications}, vol.~5, p.~3250, dec 2014.

\bibitem{Rolly2012boosting}
B.~Rolly, B.~Stout, and N.~Bonod, ``{Boosting the directivity of optical
  antennas with magnetic and electric dipolar resonant particles},'' {\em
  Optics Express}, vol.~20, p.~20376, aug 2012.

\bibitem{Hancu2014multipolar}
I.~M. Hancu, A.~G. Curto, M.~Castro-L{\'{o}}pez, M.~Kuttge, and N.~F. van
  Hulst, ``{Multipolar Interference for Directed Light Emission},'' {\em Nano
  Letters}, vol.~14, pp.~166--171, jan 2014.

\bibitem{evlyukhin2012demonstration}
A.~B. Evlyukhin, S.~M. Novikov, U.~Zywietz, R.~L. Eriksen, C.~Reinhardt, S.~I.
  Bozhevolnyi, and B.~N. Chichkov, ``{Demonstration of Magnetic Dipole
  Resonances of Dielectric Nanospheres in the Visible Region},'' {\em Nano
  Letters}, vol.~12, pp.~3749--3755, jul 2012.

\bibitem{kuznetsov2016optically}
A.~I. Kuznetsov, A.~E. Miroshnichenko, M.~L. Brongersma, Y.~S. Kivshar, and
  B.~Luk'yanchuk, ``{Optically resonant dielectric nanostructures},'' {\em
  Science}, vol.~354, p.~aag2472, nov 2016.

\bibitem{permyakov2015probing}
D.~Permyakov, I.~Sinev, D.~Markovich, P.~Ginzburg, A.~Samusev, P.~Belov,
  V.~Valuckas, A.~I. Kuznetsov, B.~S. Luk'yanchuk, A.~E. Miroshnichenko, D.~N.
  Neshev, and Y.~S. Kivshar, ``{Probing magnetic and electric optical responses
  of silicon nanoparticles},'' {\em Applied Physics Letters}, vol.~106,
  p.~171110, apr 2015.

\bibitem{wozniak2015selective}
P.~Wo{\'{z}}niak, P.~Banzer, and G.~Leuchs, ``{Selective switching of
  individual multipole resonances in single dielectric nanoparticles},'' {\em
  Laser {\&} Photonics Reviews}, vol.~9, pp.~231--240, mar 2015.

\bibitem{zywietz2014laser}
U.~Zywietz, A.~B. Evlyukhin, C.~Reinhardt, and B.~N. Chichkov, ``{Laser
  printing of silicon nanoparticles with resonant optical electric and magnetic
  responses},'' {\em Nature Communications}, vol.~5, 2014.

\bibitem{picardi2019experimental}
M.~F. Picardi, M.~Neugebauer, J.~S. Eismann, G.~Leuchs, P.~Banzer, F.~J.
  Rodr{\'{i}}guez-Fortu{\~{n}}o, and A.~V. Zayats, ``{Experimental
  demonstration of linear and spinning Janus dipoles for polarisation and
  wavelength selective near-field coupling},'' {\em Light: Science \&
  Applications}, no.~8, p.~52, 2019.

\bibitem{picardi2018not}
M.~F. Picardi, A.~V. Zayats, and F.~J. Rodr{\'{i}}guez-Fortu{\~{n}}o, ``{Not
  every dipole is the same: the hidden patterns of dipolar near fields},'' {\em
  Europhysics News}, vol.~49, pp.~14--18, jul 2018.

\bibitem{rodriguez2013near}
F.~J. Rodr{\'{i}}guez-Fortu{\~{n}}o, G.~Marino, P.~Ginzburg, D.~O'Connor,
  A.~Mart{\'{i}}nez, G.~A. Wurtz, and A.~V. Zayats, ``{Near-field interference
  for the unidirectional excitation of electromagnetic guided modes.},'' {\em
  Science}, vol.~340, no.~6130, pp.~328--30, 2013.

\bibitem{petersen2014chiral}
J.~Petersen, J.~Volz, and A.~Rauschenbeutel, ``{Chiral nanophotonic waveguide
  interface based on spin-orbit interaction of light},'' {\em Science},
  vol.~346, pp.~67--71, oct 2014.

\bibitem{mitsch2014quantum}
R.~Mitsch, C.~Sayrin, B.~Albrecht, P.~Schneeweiss, and A.~Rauschenbeutel,
  ``{Quantum state-controlled directional spontaneous emission of photons into
  a nanophotonic waveguide},'' {\em Nature Communications}, vol.~5, p.~5713,
  dec 2014.

\bibitem{kapitanova2014photonic}
P.~V. Kapitanova, P.~Ginzburg, F.~J. Rodr{\'{i}}guez-Fortu{\~{n}}o, D.~S.
  Filonov, P.~M. Voroshilov, P.~A. Belov, A.~N. Poddubny, Y.~S. Kivshar, G.~A.
  Wurtz, and A.~V. Zayats, ``{Photonic spin Hall effect in hyperbolic
  metamaterials for polarization-controlled routing of subwavelength modes},''
  {\em Nature Communications}, vol.~5, p.~3226, dec 2014.

\bibitem{neugebauer2014polarization}
M.~Neugebauer, T.~Bauer, P.~Banzer, and G.~Leuchs, ``{Polarization tailored
  light driven directional optical nanobeacon.},'' {\em Nano letters}, vol.~14,
  pp.~2546--2551, may 2014.

\bibitem{OConnor2014spin}
D.~O'Connor, P.~Ginzburg, F.~J. Rodr{\'{i}}guez-Fortu{\~{n}}o, G.~A. Wurtz, and
  A.~V. Zayats, ``{Spin–orbit coupling in surface plasmon scattering by
  nanostructures},'' {\em Nature Communications}, vol.~5, p.~5327, nov 2014.

\bibitem{coles2016chirality}
R.~Coles, D.~Price, J.~Dixon, B.~Royall, E.~Clarke, P.~Kok, M.~Skolnick,
  A.~Fox, and M.~Makhonin, ``Chirality of nanophotonic waveguide with embedded
  quantum emitter for unidirectional spin transfer,'' {\em Nature
  communications}, vol.~7, p.~11183, 2016.

\bibitem{young2015polarization}
A.~B. Young, A.~Thijssen, D.~M. Beggs, P.~Androvitsaneas, L.~Kuipers, J.~G.
  Rarity, S.~Hughes, and R.~Oulton, ``Polarization engineering in photonic
  crystal waveguides for spin-photon entanglers,'' {\em Physical review
  letters}, vol.~115, no.~15, p.~153901, 2015.

\bibitem{luxmoore2013optical}
I.~Luxmoore, N.~Wasley, A.~Ramsay, A.~Thijssen, R.~Oulton, M.~Hugues, A.~Fox,
  and M.~Skolnick, ``Optical control of the emission direction of a quantum
  dot,'' {\em Applied Physics Letters}, vol.~103, no.~24, p.~241102, 2013.

\bibitem{bliokh2015quantum}
K.~Y. Bliokh, D.~Smirnova, and F.~Nori, ``{Quantum spin Hall effect of
  light},'' {\em Science}, vol.~348, pp.~1448--1451, jun 2015.

\bibitem{slobozhanyuk2019near}
A.~Slobozhanyuk, A.~V. Shchelokova, X.~Ni, S.~Hossein~Mousavi, D.~A. Smirnova,
  P.~A. Belov, A.~Al{\`u}, Y.~S. Kivshar, and A.~B. Khanikaev, ``Near-field
  imaging of spin-locked edge states in all-dielectric topological
  metasurfaces,'' {\em Applied Physics Letters}, vol.~114, no.~3, p.~031103,
  2019.

\bibitem{vanmechelen2016universal}
T.~{Van Mechelen} and Z.~Jacob, ``{Universal spin-momentum locking of
  evanescent waves},'' {\em Optica}, vol.~3, p.~118, feb 2016.

\bibitem{picardi2018janus}
M.~F. Picardi, A.~V. Zayats, and F.~J. Rodr{\'{i}}guez-Fortu{\~{n}}o, ``{Janus
  and Huygens Dipoles: Near-Field Directionality Beyond Spin-Momentum
  Locking},'' {\em Physical Review Letters}, vol.~120, p.~117402, mar 2018.

\bibitem{picardi2017unidirectional}
M.~F. Picardi, A.~Manjavacas, A.~V. Zayats, and F.~J.
  Rodr{\'{i}}guez-Fortu{\~{n}}o, ``{Unidirectional evanescent-wave coupling
  from circularly polarized electric and magnetic dipoles: An angular spectrum
  approach},'' {\em Physical Review B}, vol.~95, p.~245416, jun 2017.

\bibitem{mandel1995optical}
L.~Mandel and E.~Wolf, {\em {Optical coherence and quantum optics}}.
\newblock Cambridge university press, 1995.

\bibitem{novotny2012principles}
L.~Novotny and B.~Hecht, {\em {Principles of Nano-Optics}}.
\newblock Cambridge: Cambridge University Press, 2012.

\bibitem{nieto2006scattering}
M.~Nieto-Vesperinas, {\em {Scattering and Diffraction in Physical Optics}}.
\newblock WORLD SCIENTIFIC, jun 2006.

\bibitem{balanis2005antenna}
C.~A. Balanis, {\em {Antenna theory : analysis and design}}.
\newblock Wiley Interscience, 2005.

\bibitem{picardi2019calculator}
M.~F. Picardi, A.~V. Zayats, and F.~J. Rodr{\'{i}}guez-Fortu{\~{n}}o,
  ``{Angular spectrum engineering of dipolar sources},'' {\em Code available on
  Zenodo at the link https://doi.org/10.5281/zenodo.3239121}, may 2019.

\bibitem{neugebauer2016polarization}
M.~Neugebauer, P.~Wo{\'{z}}niak, A.~Bag, G.~Leuchs, and P.~Banzer,
  ``{Polarization-controlled directional scattering for nanoscopic position
  sensing},'' {\em Nature Communications}, vol.~7, p.~11286, apr 2016.

\bibitem{Wei2016excitation}
L.~Wei, Z.~Xi, N.~Bhattacharya, and H.~P. Urbach, ``{Excitation of the
  radiationless anapole mode},'' {\em Optica}, vol.~3, p.~799, aug 2016.

\end{thebibliography}
\end{document}